\documentclass[12pt]{article}
\usepackage[a4paper]{geometry}
\usepackage{amsmath}
\usepackage{graphicx}
\usepackage[modulo]{lineno}
\usepackage{hyperref}
\usepackage[all]{hypcap}
\usepackage[numbers,round,compress]{natbib}
\usepackage{fancyhdr}

\newcommand{\overbar}[1]{\mkern 1.5mu\overline{\mkern-1.5mu#1\mkern-1.5mu}\mkern 1.5mu}

\title{Revealing the cluster of slow transients behind a large slow slip event}

\author{William B. Frank$^{1}$\footnote{Now at Department of Earth Sciences, University of Southern California, Los Angeles, CA, USA}\protect\phantom{\footnotesize 1}\footnote{Corresponding author (e-mail: \texttt{wbfrank@usc.edu})} , Baptiste Rousset$^{2}$, C\'{e}cile Lasserre$^{3}$,\\ and Michel Campillo$^{1,4}$}

\date{}

\begin{document}
\maketitle

\noindent$^{1}$Department of Earth, Atmospheric, and Planetary Sciences, Massachusetts Institute of Technology, Cambridge, MA, USA

\noindent$^{2}$Department of Earth and Planetary Sciences, University of California, Berkeley, CA, USA

\noindent$^{3}$Laboratoire de G\'{e}ologie de Lyon: Terre, Plan\`{e}tes, Environnement, Universit\'{e} de Lyon, Ecole Normale Sup\'{e}rieure de Lyon, Villeurbanne, France

\noindent $^{4}$Institut des Sciences de la Terre, Universit\'{e} Grenoble Alpes, CNRS, IRD, Saint-Martin-d'H\`{e}res, France

%TC:break Abstract
\vspace{0.5cm}
\textbf{
Capable of reaching similar magnitudes to large megathrust earthquakes ($M_w>7$), slow slip events play a major role in accommodating tectonic motion on plate boundaries through predominantly aseismic rupture.
We demonstrate here that large slow slip events are a cluster of short-duration slow transients.
Using a dense catalog of low-frequency earthquakes as a guide, we investigate the $M_w7.5$ slow slip event that occurred in 2006 along the subduction interface 40~km beneath Guerrero, Mexico.
We show that while the long-period surface displacement as recorded by GPS suggests a six month duration, motion in the direction of tectonic release only sporadically occurs over 55 days and its surface signature is attenuated by rapid relocking of the plate interface.
Our proposed description of slow slip as a cluster of slow transients forces us to reevaluate our understanding of the physics and scaling of slow earthquakes.
}
\vspace{0.5cm}

%TC:break Main text
\clearpage
\baselineskip=2\baselineskip % double spacing
%\linenumbers
\pagestyle{fancy}

\section*{Introduction}
Slow slip events~\cite{Dragert:2001}, like other slow earthquakes~\cite{Beroza:2011} such as tectonic tremor~\cite{Obara:2002} and low-frequency earthquakes (LFEs)~\cite{Shelly:2006}, occur downdip of the seismogenic zone where increasing temperatures and pressures transition the faulting style from brittle stick-slip towards stable sliding~\cite{Hyndman:1993,Scholz:2002,Avouac:2015}.
High pore fluid pressures maintained by the metamorphic dehydration of the downgoing slab impose small stress drops on any events that nucleate within this region and potentially inhibit fast rupture~\cite{Shelly:2006,Liu:2007,Audet:2009,Segall:2010}.
Given that the seismic moment of tremors and LFEs is negligible with respect to the geodetic moment of slow earthquakes~\cite{Kostoglodov:2010,Maury:2018}, slow slip events are primarily observed with continuous GPS measurements at the surface~\cite{Nishimura:2014,Rousset:2017} whose temporal resolution is often limited to daily position solutions~\cite{Herring:2016}.
Current numerical models constrained with such geodetic observations suggest that slow slip is the long-duration, steady rupture of the aseismic matrix on the subduction interface~\cite{Liu:2007,Radiguet:2011,Perfettini:2008}.
The seismic asperities embedded within the aseismic fault material, which are responsible for tectonic tremor and LFEs, are transiently loaded by slow slip, resulting in accelerated seismicity rates~\cite{Frank:2015a}.
Recent studies have shown it is possible to use these seismic crackles and pops to directly geodetically observe the underlying slow deformation~\cite{Frank:2015b,Frank:2016c}.

The subhorizontal subduction zone beneath Guerrero, Mexico shown in Fig.~\ref{decomposition}a hosts a slow slip cycle that releases every four years more accumulated tectonic strain than a $M_w$~7 earthquake~\cite{Radiguet:2012}.
Focusing on one of the most studied instances of this cycle, the continuous GPS displacement time series in Fig.~\ref{decomposition}b highlights a six-month $M_w$~7.5 slow slip event in 2006.
A geodetic kinematic model of this slow slip event reproduces the surface observations with a smooth rupture that lasts 185~days, accumulating more than 15~cm of slip on the plate interface~\cite{Radiguet:2011}.
While the modeled slip history reproduces the long-period surface displacements as recorded by GPS, recent work has highlighted that there is coherent information at shorter time-scales within the GPS time series that can be extracted using LFE/tremor activity as a guide~\cite{Frank:2016c}.
In this context, we perform a multidisciplinary investigation of the fine-scale evolution of the 2006 slow slip event using a dense catalog of LFEs~\cite{Frank:2014b}.

\section*{Results}
\subsection*{Decomposition of surface motion via low-frequency earthquakes}
Guerrero LFEs occur in two different source regions~\cite{Frank:2014b} (see Fig.~\ref{decomposition}).
The sweet spot that is located furthest downdip exhibits a near continuous stream of event bursts, each burst thought to coincide with a small slip event~\cite{Frank:2013}.
In the transient zone, closer to the trench within the main slow slip source region, there is a strong correlation between LFEs and geodetically observed slow slip events~\cite{Frank:2015a}.
We therefore focus on the 34,389 LFEs spread over the 58 repeating sources in the transient zone to geodetically investigate the 2006 slow slip event.

We use this LFE catalog to decompose the surface displacement time series during the 2006 slow slip event~~\cite{Radiguet:2011} recorded at five GPS stations that lie between Acapulco and Mexico City.
We first define the daily LFE activity in the transient zone as the product of the daily number of LFEs and the daily median LFE seismic amplitude, which we call the daily LFE amplitude sum time series (see Methods).
Considering high LFE activity in the transient zone as an indicator of when the subduction interface is slowly slipping~\cite{Frank:2015b,Frank:2016c}, we then threshold the daily LFE amplitude sum time series in Fig.~\ref{decomposition}c to determine whether each daily N-S GPS displacement increment should represent tectonic release (slow slip) or loading (plate locking).
Tectonic loading is represented by surface motion towards the North when the subducting Cocos plate is locked underneath the North American plate, while tectonic release corresponds to motion towards the South when the subduction interface decouples and built-up tectonic stress is released.
By defining the daily LFE amplitude sum to include all transient zone LFE activity, we sacrifice spatial resolution to increase the temporal resolution of our analysis.
This compromise allows us to geodetically detect slow slip on the same time scale as the sampling rate of the GPS time series.

\subsection*{Intermittent and clustered evolution of slow slip}
\begin{figure}
\makebox[\textwidth][c]{\includegraphics{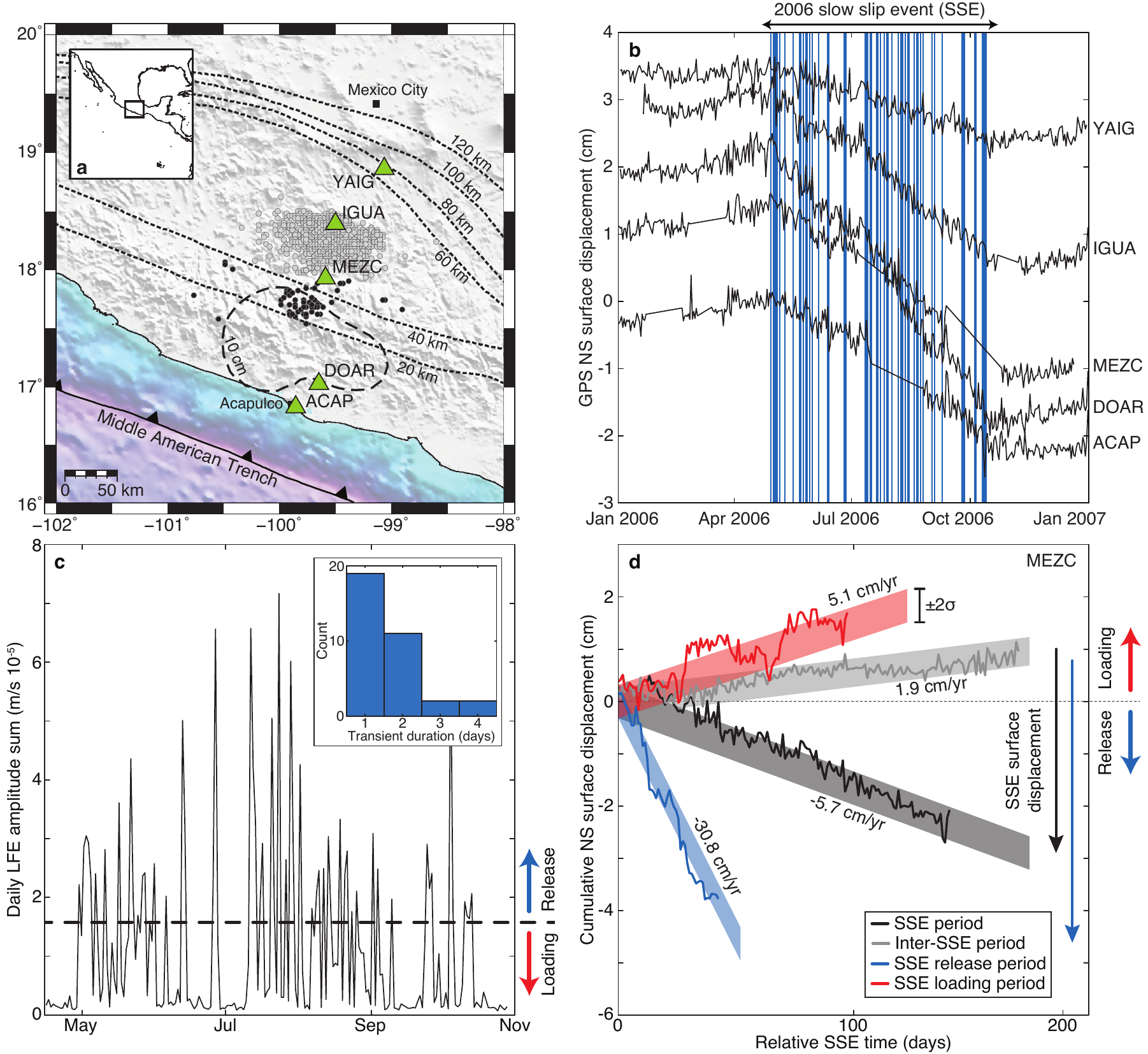}}
\caption{\footnotesize{
\textbf{Breaking down a large slow slip event into its constituent slow transients.}
(a)~Tectonic context of the subduction zone underneath Guerrero, Mexico.
A large slow slip event in 2006 recorded by GPS (green triangles) accumulated more than 10~cm of slip (thick dashed contour) updip of low-frequency earthquake (LFE) sources~\cite{Frank:2014b} (black points: transient zone; gray points: sweet spot).
Depth contours of the subduction interface~\cite{Kim:2010} are shown as thin dashed lines.
(b)~GPS displacement time series during 2006.
The blue patches indicate the set of slow transients that exhibit tectonic release.
(c)~Daily LFE amplitude sums during the 2006 slow slip event.
We identify slow transients on the days that the daily amplitude sum exceeds the established threshold (dashed black line).
The insert shows the distribution of their durations.
(d)~Cumulative displacements at MEZC during the 2006 slow slip event before and after decomposition via the LFE amplitude sum in (c).
The black trace represents the displacements during the 185~day slow slip duration~\cite{Radiguet:2011} in (b), while the gray trace shows the inter-SSE (Slow Slip Event) displacements during the 185~days before the 2006 event.
The red and blue traces respectively show the decomposed loading and release displacements (see text).
The shaded regions represent the estimated motion $\pm2\sigma$ during the slow slip duration of 185~days, of which there is no data at MEZC for 35~days.
The slow slip induced surface displacement of the cluster of slow transients during the release period of the 2006 slow slip event (blue arrow) is 40\% larger than the surface displacement estimated from from the GPS time series during the full slow slip duration (black arrow).
}}
\label{decomposition}
\end{figure}

The decomposition shown in Fig.~\ref{decomposition}d of MEZC, the GPS station directly above the analyzed LFE activity and most sensitive to the slow slip in the vicinity of the transient zone, demonstrates that there are both loading and release regimes mixed together within the noisy surface displacements.
We find not only a greater release displacement than the long-period signature that lasts three times longer, but also northward loading during what was previously considered to be continuous slip.
Once episodes of tectonic release are highlighted, visually striking periods of northward loading are evident in the GPS time series, such as the one in July 2006 (Fig.~\ref{decomposition}b).
We observe this same separation of surface motion into two distinct regimes of release and loading on all of the analyzed GPS stations (see Materials and Methods and figs.~S1 and S2).
Considering continuous days of release as a single slow transient, we find that the average duration of slip is about 1--2~days (Fig.~\ref{decomposition}c).
Taking into account the scattered temporal distribution of tectonic release over 55~days seen in Fig.~\ref{decomposition}b, this implies that slow transient slip is not continuous and is interrupted by intermittent locking of the plate interface.
From the estimated loading velocity of 5.1~cm/yr at MEZC that is only slightly lower than the plate convergence rate (6.4~cm/yr), we infer that the subduction interface is at times completely locked during slow slip.
This is also reported at multiple plate boundaries during the inter-SSE (Slow Slip Event) phase of the slow slip cycle~\cite{Frank:2016c}, defined as the time period between large slow slip events.
Such work shows that long-term loading rates are biased by intermittent release and locking that reveals strong plate coupling over short time scales.

\begin{figure}
\makebox[\textwidth][c]{\includegraphics{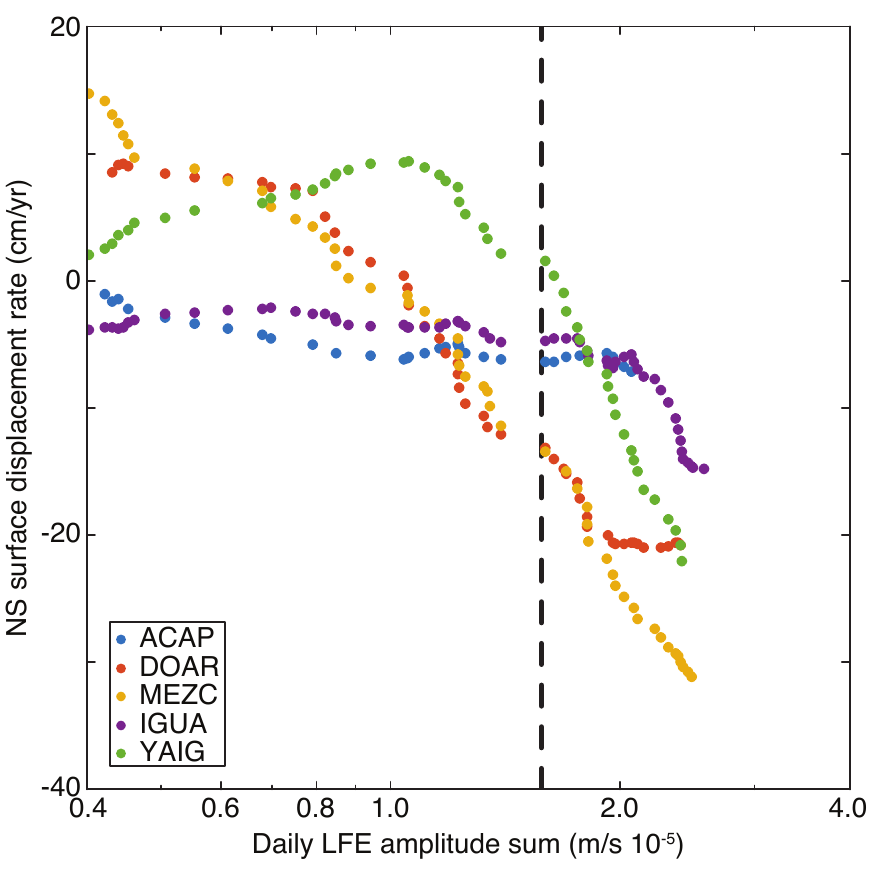}}
\caption{\textbf{Estimating surface displacement rates via low-frequency earthquake (LFEs) amplitudes on the subduction interface.}
Surface displacement rates are computed at each GPS station as a function of increasing daily LFE amplitude sums.
Southward surface motion in the direction of tectonic release, which is proportional to the motion on the decoupled interface at depth, becomes pronounced at LFE amplitudes greater than the established threshold (dashed line).
}
\label{displacement_rate}
\end{figure}

To home in on the fine-scale behavior of the slowly slipping plate interface, we compute the cumulative displacements at each GPS station as a function of increasing daily LFE amplitude sums, regardless of the tectonic regime.
The smoothed slope of these cumulative displacements highlights the strong dependence of the surface displacement rate on LFE activity as shown in Fig.~\ref{displacement_rate}.
Because the surface displacement rate is proportional to the slip rate on the subhorizontal plate interface in Guerrero, we suggest that the evolution of the slip rate during slow slip mirrors the observed complex time-history of the low-frequency seismicity (Fig.~\ref{decomposition}c).
This is in contrast to previous theoretical~\cite{Liu:2007,Perfettini:2008} and data-driven~\cite{Radiguet:2011} models of smooth large-scale slow slip events and suggests the complex time history of slow slip drives the intricate patterns of slow seismicity that are reported~\cite{Shelly:2007,Houston:2011,Hawthorne:2013,Peng:2015,Frank:2016a}.

\begin{figure}
\makebox[\textwidth][c]{\includegraphics{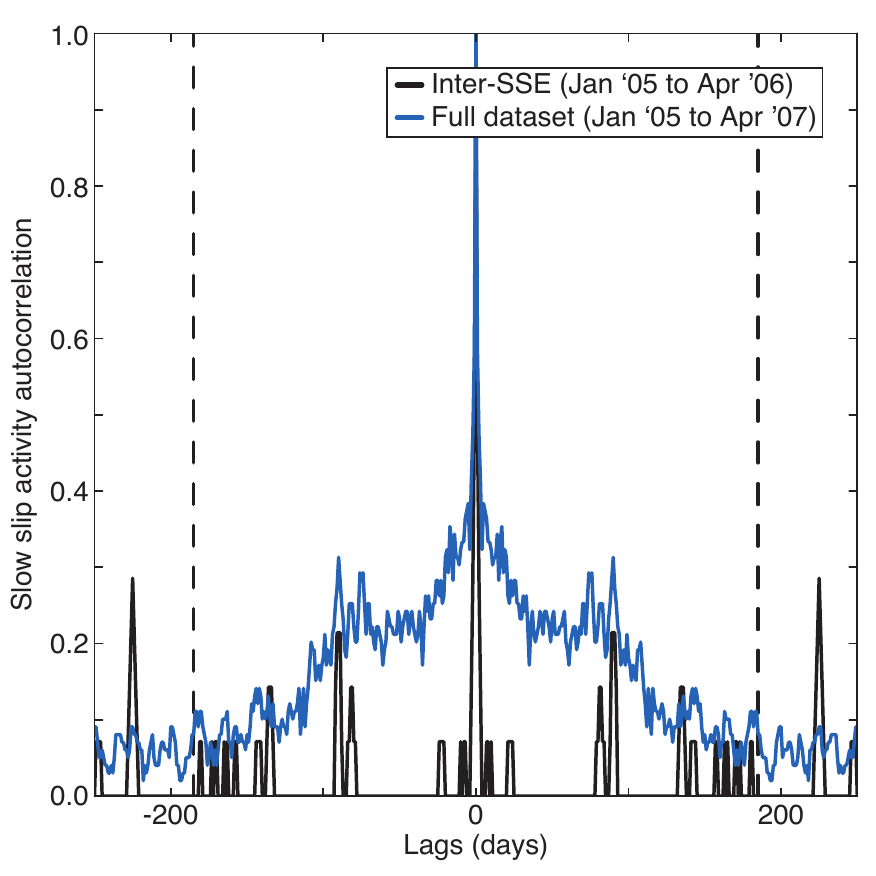}}
\caption{\textbf{Slow slip as a cluster of slow transients.}
The autocorrelation of the slow slip activity time series in blue (see text) indicates a temporally clustered distribution of slow transients with a falloff of 185 days (dashed lines), the duration of the 2006 slow slip event~\cite{Radiguet:2011}.
The inter-SSE (Slow Slip Event) time period before the 2006 slow slip event (black) exhibits a Dirac at zero lag, indicative of a random occurrence of slow transients.
}
\label{sse_clustering}
\end{figure}

The intermittent slow deformation observed here is reminiscent of reports of temporally clustered LFE activity~\cite{Frank:2016a,Lengline:2017}.
To evaluate whether the timing of the slow transients mimics the clustered distribution of LFE activity, we analyze the two-and-a-half year duration of the LFE catalog from Jan~1~2005 to Apr~15~2007.
We first divide the previous LFE amplitude sum threshold by two to account for the lower LFE rates before and after the 2006 event~\cite{Frank:2015a}.
We then generate a regularly-sampled binary slow slip activity time series: 1 on days when the daily LFE amplitude sum exceeds the threshold and slow slip is considered to occur, and 0 for every other day.
The autocorrelation of this time series in Fig.~\ref{sse_clustering} shows a smooth falloff from zero lag that indicates that the timing of the slow transients is not random and their occurrence is clustered~\cite{Frank:2016a}.
We also observe that the clustering falls off until 185~days, which corresponds to the long-period duration of the 2006 slow slip event.
This demonstrates that the observed intermittent tectonic release during the large slow slip event represents a clustered occurrence of slow transients, where each slow transient triggers and is triggered by other slip events.
If we analyze only the time period before the 2006 slow slip event, we see that this clustering disappears and is replaced by a delta function at zero lag, implying a random temporal distribution of slow slip prior to the large 2006 event~\cite{Frank:2016a}.

\section*{Discussion}
\subsection*{Redefining slow slip as a cluster of slow transients}
A new description of slow slip emerges from the set of observations presented here: once the subduction interface decouples, it provokes a cluster of short-duration slow transients that can last for several months.
Comparing the average slow transient duration of 1--2~days that we observe to the six-month long-period signal~\cite{Radiguet:2011} in Fig.~\ref{decomposition}, we suggest that each slow transient is a short-duration slip pulse.
Where the conditions are right for tremors and LFEs~\cite{Shelly:2006,Audet:2009}, each individual slip pulse briefly increases the stressing rate on nearby asperities, triggering a burst of seismic activity~\cite{Frank:2016a} before the subduction interface relocks.
Surface geodetic observations at long time-scales are unable to capture this fine-scale activity and only sample  the smooth envelope of a cluster of slow transients.
Our conceptual model thus accounts for both the aseismic and seismic observables of large slow slip events.
This description of a large slow slip event is also consistent with observations earlier in the slow slip cycle, where intermittent release and locking define the inter-SSE period in between slow slip events~\cite{Frank:2016c}.

To produce the observed highly variable slip rate and clustering behavior, we suggest that the frictional heterogeneity of the subduction interface is the dominant factor in controlling the evolution of slow slip~\cite{BenZion:2008}.
This implies that the heterogeneity that governs the ruptures of megathrust earthquakes in the seismogenic zone~\cite{Zhao:2011,Aochi:2011} is preserved during subduction and plays a major role in how tectonic motion is accommodated at greater depths.
Previous numerical work~\cite{Nakata:2011,Ando:2012,Skarbek:2012,Dublanchet:2013,Yabe:2017} has also suggested that the frictional heterogeneity along a fault can reproduce many of the observables of slow earthquakes, with a complex evolution of slip on brittle asperities controlled by slow aseismic slip in the surrounding fault material.
Constant background aseismic slip as the driving mechanism behind slow slip is not consistent, however, with the significant periods of tectonic loading that we observe during slow slip (Fig.~\ref{decomposition}).
Our results negate the possibility of such a large-scale slow aseismic slip front that would link the individual slow transients into a cluster, because the observed loading rates imply a locked plate interface.
Any potential mechanism behind the clustered slow transients we observe here would have to be able to govern the interaction between slow transients along a locked fault.
One such mechanism for which there is abundant geological evidence~\cite{Fagereng:2011,Fisher:2014,Audet:2014,Taetz:2018} is the rapid diffusion of high pore fluid pressures during faulting at depth~\cite{Frank:2015a,Rice:1992,Skarbek:2016}.

Another significant consequence of the intermittent relocking we observe during slow slip is the attenuation of the surface motion as recorded by GPS.
This accounts for the 40\% larger surface displacements that we observe during the release regime in Fig.~\ref{decomposition}d compared to the long-period geodetic estimates~\cite{Radiguet:2011}.
We consequently infer that the long-period measurements of surface displacement that inform previous models of large slow slip events systematically underestimate their moment magnitude.
We note there is the possibility that this bias could also affect previously determined source locations of large slow slip events.
This intermittent locking likely depends on the dominant style of faulting that varies with depth~\cite{Hyndman:1993,Scholz:2002}, implying a moment underestimation that varies along the plate interface with distance from the trench.
This would impact the distribution of surface displacements during slow slip that inform geodetic fault slip inversions, consequently impacting the inferred source location of slow slip.

\subsection*{Conclusions}
By breaking down a large slow slip event into a cluster of slow transients, we demonstrate that previous studies of large slow slip events both overestimate their duration $T$ and underestimate their moment magnitude $M$.
Our multidisciplinary analysis of the 2006 slow slip event yields a three-times shorter duration, and assuming the same spatial distribution of slip as previous studies~\cite{Radiguet:2011}, a moment that is at least 40\% larger than the previous geodetic estimate.
If we impose a similarly shorter duration and larger moment on all large slow slip events observed at plate boundaries, the proposed $M\sim T$ slow earthquake scaling~\cite{Ide:2007} will not shift directly to a classical $M\sim T^3$ earthquake scaling; it will instead likely satisfy a scaling relationship with an exponent between 1 and 2 that is consistent with fractal distributions of fault slip~\cite{Ide:2008,Ide:2010,BenZion:2012}.
This bias we observe will, however, displace all of the observations of large slow slip events that constrain the proposed $M\sim T$ slow earthquake scaling~\cite{Ide:2007} at long durations and large moments.
Another possible interpretation is that each of the short-duration slow transients should be characterized as separate slow earthquakes.
This ignores, however, the characteristic clustering signature~\cite{Frank:2016a} that links temporally disparate slow transients together to create a large slow slip event.
In any case, our results contribute to a growing body of evidence~\cite{Bostock:2015,Gomberg:2016} that we must reevaluate our understanding of the physics and scaling of slow earthquakes in light of new observational constraints.

%TC:break Methods
\section*{Materials and Methods}
\subsection*{Daily low-frequency earthquake amplitude sums}
We compute the daily amplitude sum of LFEs as follows:
\begin{equation}
\overbar{A}(t) = N(t) A(t),
\end{equation}
where $t$ is time (in days), $N(t)$ is the daily count of LFEs, and $A(t)$ is the daily median amplitude of the cataloged LFEs~\cite{Frank:2014b}.
The daily median LFE amplitude $A(t)$ is determined as:
\begin{equation}
A(t) = \underset{i}{\operatorname{median}} \left[ \underset{s}{\operatorname{median}} \left( \frac{\Sigma_c^C a_i^{s,c}}{C} \right) \right],
\end{equation}
where $a_i$ represents the peak amplitude of the $i^{\textrm{th}}$ LFE on a given day $t$, $s$ represents the ten seismic stations used to generate the LFE catalog, and $c$ represents the $C=3$ components.

\subsection*{Decomposition of GPS displacement increments via low-frequency earthquake amplitude sums}
We first compute the daily GPS NS displacement increments ${\Delta x}_s$ at each station $s$ as:
\begin{equation}
{\Delta x}_s(t) = x_s(t + 1) - x_s(t),
\end{equation}
where $t$ is time in days and $x(t)$ is the GPS position time series.
We propagate the observational position errors $\epsilon_s(t)$ to the displacement increment errors ${\Delta \epsilon}_s(t)$  as follows:
\begin{equation}
{\Delta \epsilon}_s(t) = \sqrt{\epsilon_s(t)^2 + \epsilon_s(t + 1)^2}.
\label{observation_error}
\end{equation}

We then analyze the daily LFE amplitude sums with respect to some threshold to determine when the subduction interface is slipping and releasing tectonic stress or loading and accumulating stress.
For example, during tectonic release, there should be significant LFE activity associated with slow slip and a consequent high amplitude sum greater than the established threshold.
During the loading regime, there should be little to no activity and a low amplitude sum smaller than the threshold while the plate interface is locked and coupled.

After this sorting is completed, we have four different sets of displacement increments that are each associated with a different tectonic regime: the long-period 185-day \textit{SSE} (Slow Slip Event) duration, the \textit{inter-SSE} period, defined here as the 185 days before the 2006 slow slip event, and the \textit{release} and \textit{loading} regimes during the slow slip event as defined by LFE activity.
Because we observe a stationary distribution of displacement increments for each regime, we estimate the displacement uncertainty as the mean of the observational errors ${\Delta \epsilon}_s(t)$ for each regime.
We report this average observational error $\overbar{\Delta \epsilon}_s(t)$ multiplied by four as the $4\sigma$ width of the shaded regions in Fig.~\ref{decomposition} and fig.~S1.
To then measure the average surface velocity from a set of GPS displacement increments, we know that the mean is not stable because only a fraction of the displacement increments contribute to the estimate~\cite{Frank:2016c}.
We therefore determine the average surface velocity as the slope of the linear regression of the cumulative displacement time series as it takes into account every displacement increment datum.
Given that the displacement increments represent a relative displacement that occurs in a given regime regardless of the its time stamp, the measured velocity from the cumulative displacement time series of a random resampling of the displacement increments should be the same.
We therefore randomly resample the displacement increments (allowing for a given datum to be selected multiple times) 10,000 times for each regime and compute a cumulative displacement time series for each resampling.
We then perform a linear regression of all 10,000 iterations, with each point weighted by $\overbar{\Delta \epsilon}_s(t)^{-2}$, to determine the average velocity for each regime as the slope of the best-fit linear trend.
We then compute the surface displacement as the average velocity multiplied by the duration of that regime as defined by the daily LFE amplitude sum time series in Fig.~\ref{decomposition}c.
In such a way, we avoid any biases associated with data gaps in the geodetic time series.
As described above, we then report the uncertainty of the displacement as the average observational error.

Now that we can compute displacements and velocities for each regime given some threshold, we tested all possible thresholds to determine the best threshold that yielded the largest different displacement, defined as the loading displacement minus the release displacement.
Given that there are a finite number of daily LFE amplitude sums, we used each as a potential threshold and performed the GPS decomposition for the loading and release regimes as described above.
We then stacked the differential displacements over the five analyzed GPS stations in fig.~S3 and picked the threshold that produced the maximum stacked differential displacement.

\subsection*{Robustness of GPS displacement decomposition}
The network sum of the estimated release displacements provides a single quantity for each iteration that represents how effective the decomposition into loading and release was.
To evaluate whether the observed decomposed displacements in Fig.~\ref{decomposition} and fig.~S1 could happen by chance, we employ the following bootstrap analysis.

We randomly shuffle the time $t$ of the daily LFE amplitude sum time series without modifying the time $t$ of the GPS displacement increments and reperform the decomposition at all stations.
We perform this shuffling 10,000 times and compare the observed network release displacement to the distribution of random network release displacements in fig.~S2.
Given that the observed displacement is more than $3\sigma$ from the mean of the random network release displacements, we conclude that there is a negligible chance our observations result from a random decomposition.

%\subsection*{Data and code availability}
%The GPS time series, the Guerrero LFE catalog, and computer code used to analyze the data is available upon request to the corresponding author, W.B.F.

%\subsection*{Author contributions}
%W.B.F. performed the data analysis.
%All authors contributed to interpreting the results and writing the manuscript.

\section*{Supplementary Materials}
\begin{itemize}
  \item fig.~S1: Decomposition of surface displacement increments into loading and release in Guerrero, Mexico.
  \item fig.~S2: Comparing observed network release displacement to a random shuffling of GPS displacement increments.
  \item fig.~S3: Determining the low-frequency earthquake amplitude sum threshold.
\end{itemize}

%:BIBLIOGRAPHY
\bibliographystyle{unsrtnat}
%\bibliography{decompose_sse}

\begin{thebibliography}{50}
\providecommand{\natexlab}[1]{#1}
\providecommand{\url}[1]{\texttt{#1}}
\expandafter\ifx\csname urlstyle\endcsname\relax
  \providecommand{\doi}[1]{doi: #1}\else
  \providecommand{\doi}{doi: \begingroup \urlstyle{rm}\Url}\fi

\bibitem[Dragert et~al.(2001)Dragert, Wang, and James]{Dragert:2001}
Herb Dragert, Kelin Wang, and Thomas~S James.
\newblock {A silent slip event on the deeper Cascadia subduction interface}.
\newblock \emph{Science}, 292\penalty0 (5521):\penalty0 1525--1528, 2001.

\bibitem[Beroza and Ide(2011)]{Beroza:2011}
Gregory~C Beroza and Satoshi Ide.
\newblock {Slow earthquakes and nonvolcanic tremor}.
\newblock \emph{Annu. Rev. Earth Planet. Sci.}, 39:\penalty0 271--296, 2011.

\bibitem[Obara(2002)]{Obara:2002}
Kazushige Obara.
\newblock {Nonvolcanic deep tremor associated with subduction in southwest
  Japan}.
\newblock \emph{Science}, 296\penalty0 (5573):\penalty0 1679--1681, 2002.

\bibitem[Shelly et~al.(2006)Shelly, Beroza, Ide, and Nakamula]{Shelly:2006}
David~R Shelly, Gregory~C Beroza, Satoshi Ide, and Sho Nakamula.
\newblock {Low-frequency earthquakes in Shikoku, Japan, and their relationship
  to episodic tremor and slip}.
\newblock \emph{Nature}, 442\penalty0 (7099):\penalty0 188--191, 2006.

\bibitem[Hyndman and Wang(1993)]{Hyndman:1993}
Robert~D Hyndman and Kelin Wang.
\newblock {Thermal constraints on the zone of major thrust earthquake failure:
  The Cascadia subduction zone}.
\newblock \emph{J. Geophys. Res.}, 98\penalty0 (B2):\penalty0 2039--2060, 1993.

\bibitem[Scholz(2002)]{Scholz:2002}
Christopher~H Scholz.
\newblock \emph{{The mechanics of earthquakes and faulting}}.
\newblock Cambridge University Press, 2002.

\bibitem[Avouac(2015)]{Avouac:2015}
Jean-Philippe Avouac.
\newblock {From geodetic imaging of seismic and aseismic fault slip to dynamic
  modeling of the seismic cycle}.
\newblock \emph{Annu. Rev. Earth Planet. Sci.}, 43:\penalty0 233--271, 2015.

\bibitem[Liu and Rice(2007)]{Liu:2007}
Yajing Liu and James~R Rice.
\newblock {Spontaneous and triggered aseismic deformation transients in a
  subduction fault model}.
\newblock \emph{J. Geophys. Res.}, 112\penalty0 (B9), 2007.

\bibitem[Audet et~al.(2009)Audet, Bostock, Christensen, and
  Peacock]{Audet:2009}
Pascal Audet, Michael~G Bostock, Nikolas~I Christensen, and Simon~M Peacock.
\newblock {Seismic evidence for overpressured subducted oceanic crust and
  megathrust fault sealing}.
\newblock \emph{Nature}, 457\penalty0 (7225):\penalty0 76--78, 2009.

\bibitem[Segall et~al.(2010)Segall, Rubin, Bradley, and Rice]{Segall:2010}
Paul Segall, Allan~M Rubin, Andrew~M Bradley, and James~R Rice.
\newblock {Dilatant strengthening as a mechanism for slow slip events}.
\newblock \emph{J. Geophys. Res.}, 115\penalty0 (B12), 2010.

\bibitem[Kostoglodov et~al.(2010)Kostoglodov, Husker, Shapiro, Payero,
  Campillo, Cotte, and Clayton]{Kostoglodov:2010}
Vladimir Kostoglodov, Allen Husker, Nikolai~M Shapiro, Juan~S Payero, Michel
  Campillo, Nathalie Cotte, and Robert Clayton.
\newblock {The 2006 slow slip event and nonvolcanic tremor in the Mexican
  subduction zone}.
\newblock \emph{Geophys. Res. Lett.}, 37\penalty0 (24), 2010.

\bibitem[Maury et~al.(2018)Maury, Ide, Cruz-Atienza, and
  Kostoglodov]{Maury:2018}
Julie Maury, Satoshi Ide, V~M Cruz-Atienza, and V~Kostoglodov.
\newblock {Spatiotemporal variations in slow earthquakes along the Mexican
  subduction zone}.
\newblock \emph{J. Geophys. Res.}, 123\penalty0 (2):\penalty0 1559--1575, 2018.

\bibitem[Nishimura et~al.(2014)Nishimura, Sato, and Sagiya]{Nishimura:2014}
Takuya Nishimura, Mariko Sato, and Takeshi Sagiya.
\newblock {Global Positioning System (GPS) and GPS-acoustic observations:
  Insight into slip along the subduction zones around Japan}.
\newblock \emph{Annu. Rev. Earth Planet. Sci.}, 42:\penalty0 653--674, 2014.

\bibitem[Rousset et~al.()Rousset, Campillo, Lasserre, Frank, Cotte,
  Walpersdorf, Socquet, and Kostoglodov]{Rousset:2017}
B.~Rousset, M.~Campillo, C.~Lasserre, W.~B. Frank, N.~Cotte, A.~Walpersdorf,
  A.~Socquet, and V.~Kostoglodov.
\newblock {A geodetic matched-filter search for slow slip with application to
  the Mexico subduction zone}.
\newblock \emph{J. Geophys. Res.}
\newblock 2017JB014448.

\bibitem[Herring et~al.(2016)Herring, Melbourne, Murray, Floyd, Szeliga, King,
  Phillips, Puskas, Santillan, and Wang]{Herring:2016}
Thomas~A Herring, Timothy~I Melbourne, Mark~H Murray, Michael~A Floyd, Walter~M
  Szeliga, Robert~W King, David~A Phillips, Christine~M Puskas, Marcelo
  Santillan, and Lei Wang.
\newblock {Plate Boundary Observatory and Related Networks: GPS Data Analysis
  Methods and Geodetic Products}.
\newblock \emph{Rev. Geophys.}, 2016.

\bibitem[Radiguet et~al.(2011)Radiguet, Cotton, Vergnolle, Campillo, Valette,
  Kostoglodov, and Cotte]{Radiguet:2011}
Mathilde Radiguet, Fabrice Cotton, Mathilde Vergnolle, Michel Campillo, Bernard
  Valette, Vladimir Kostoglodov, and Nathalie Cotte.
\newblock {Spatial and temporal evolution of a long term slow slip event: the
  2006 Guerrero Slow Slip Event}.
\newblock \emph{Geophys. J. Int.}, 184\penalty0 (2):\penalty0 816--828, 2011.

\bibitem[Perfettini and Ampuero(2008)]{Perfettini:2008}
Hugo Perfettini and Jean-Paul Ampuero.
\newblock {Dynamics of a velocity strengthening fault region: Implications for
  slow earthquakes and postseismic slip}.
\newblock \emph{J. Geophys. Res.}, 113\penalty0 (B9), 2008.

\bibitem[Frank et~al.(2015{\natexlab{a}})Frank, Shapiro, Husker, Kostoglodov,
  Bhat, and Campillo]{Frank:2015a}
William~B Frank, Nikola\"{i}~M Shapiro, Allen~L Husker, Vladimir Kostoglodov,
  Harsha~S Bhat, and Michel Campillo.
\newblock {Along-fault pore-pressure evolution during a slow-slip event in
  Guerrero, Mexico}.
\newblock \emph{Earth Planet. Sci. Lett.}, 413:\penalty0 135--143,
  2015{\natexlab{a}}.

\bibitem[Frank et~al.(2015{\natexlab{b}})Frank, Radiguet, Rousset, Shapiro,
  Husker, Kostoglodov, Cotte, and Campillo]{Frank:2015b}
William~B Frank, Mathilde Radiguet, Baptiste Rousset, Nikola{\"\i}~M Shapiro,
  Allen~L Husker, Vladimir Kostoglodov, Nathalie Cotte, and Michel Campillo.
\newblock {Uncovering the geodetic signature of silent slip through repeating
  earthquakes}.
\newblock \emph{Geophys. Res. Lett.}, 42\penalty0 (8):\penalty0 2774--2779,
  2015{\natexlab{b}}.

\bibitem[Frank(2016)]{Frank:2016c}
William~B Frank.
\newblock {Slow slip hidden in the noise: The intermittence of tectonic
  release}.
\newblock \emph{Geophys. Res. Lett.}, 43\penalty0 (19), 2016.

\bibitem[Radiguet et~al.(2012)Radiguet, Cotton, Vergnolle, Campillo,
  Walpersdorf, Cotte, and Kostoglodov]{Radiguet:2012}
Mathilde Radiguet, Fabrice Cotton, Mathilde Vergnolle, Michel Campillo, Andrea
  Walpersdorf, Nathalie Cotte, and Vladimir Kostoglodov.
\newblock {Slow slip events and strain accumulation in the Guerrero gap,
  Mexico}.
\newblock \emph{J. Geophys. Res.}, 117\penalty0 (B4), 2012.

\bibitem[Frank et~al.(2014)Frank, Shapiro, Husker, Kostoglodov, Romanenko, and
  Campillo]{Frank:2014b}
William~B Frank, Nikola{\"\i}~M Shapiro, Allen~L Husker, Vladimir Kostoglodov,
  Alexey Romanenko, and Michel Campillo.
\newblock {Using systematically characterized low-frequency earthquakes as a
  fault probe in Guerrero, Mexico}.
\newblock \emph{J. Geophys. Res.}, 119\penalty0 (10):\penalty0 7686--7700,
  2014.

\bibitem[Frank et~al.(2013)Frank, Shapiro, Kostoglodov, Husker, Campillo,
  Payero, and Prieto]{Frank:2013}
William~B Frank, Nikola{\"\i}~M Shapiro, Vladimir Kostoglodov, Allen~L Husker,
  Michel Campillo, Juan~S Payero, and Germ{\'a}n~A Prieto.
\newblock {Low-frequency earthquakes in the Mexican Sweet Spot}.
\newblock \emph{Geophys. Res. Lett.}, 40\penalty0 (11):\penalty0 2661--2666,
  2013.

\bibitem[Hawthorne and Rubin(2013)]{Hawthorne:2013}
Jessica~C Hawthorne and Allan~M Rubin.
\newblock {Short-time scale correlation between slow slip and tremor in
  Cascadia}.
\newblock \emph{J. Geophys. Res.}, 118\penalty0 (3):\penalty0 1316--1329, 2013.

\bibitem[Peng et~al.(2015)Peng, Rubin, Bostock, and Armbruster]{Peng:2015}
Yajun Peng, Allan~M Rubin, Michael~G Bostock, and John~G Armbruster.
\newblock {High-resolution imaging of rapid tremor migrations beneath southern
  Vancouver Island using cross-station cross correlations}.
\newblock \emph{J. Geophys. Res.}, 120\penalty0 (6):\penalty0 4317--4332, 2015.

\bibitem[Frank et~al.(2016)Frank, Shapiro, Husker, Kostoglodov, Gusev, and
  Campillo]{Frank:2016a}
William~B Frank, Nikola{\"\i}~M Shapiro, Allen~L Husker, Vladimir Kostoglodov,
  Alexander~A Gusev, and Michel Campillo.
\newblock {The evolving interaction of low-frequency earthquakes during
  transient slip}.
\newblock \emph{Sci. Adv.}, 2\penalty0 (4), 2016.

\bibitem[Shelly et~al.(2007)Shelly, Beroza, and Ide]{Shelly:2007}
David~R Shelly, Gregory~C Beroza, and Satoshi Ide.
\newblock {Complex evolution of transient slip derived from precise tremor
  locations in western Shikoku, Japan}.
\newblock \emph{Geochem., Geophys., Geosyst.}, 8\penalty0 (10), 2007.

\bibitem[Houston et~al.(2011)Houston, Delbridge, Wech, and
  Creager]{Houston:2011}
Heidi Houston, Brent~G Delbridge, Aaron~G Wech, and Kenneth~C Creager.
\newblock {Rapid tremor reversals in Cascadia generated by a weakened plate
  interface}.
\newblock \emph{Nature Geosci.}, 4\penalty0 (6):\penalty0 404, 2011.

\bibitem[Lenglin{\'e} et~al.(2017)Lenglin{\'e}, Frank, Marsan, and
  Ampuero]{Lengline:2017}
Olivier Lenglin{\'e}, WB~Frank, D~Marsan, and J-P Ampuero.
\newblock {Imbricated slip rate processes during slow slip transients imaged by
  low-frequency earthquakes}.
\newblock \emph{Earth Planet. Sci. Lett.}, 476:\penalty0 122--131, 2017.

\bibitem[Ben-Zion(2008)]{BenZion:2008}
Yehuda Ben-Zion.
\newblock {Collective behavior of earthquakes and faults: Continuum-discrete
  transitions, progressive evolutionary changes, and different dynamic
  regimes}.
\newblock \emph{Rev. Geophys.}, 46\penalty0 (4), 2008.

\bibitem[Zhao et~al.(2011)Zhao, Huang, Umino, Hasegawa, and
  Kanamori]{Zhao:2011}
Dapeng Zhao, Zhouchuan Huang, Norihito Umino, Akira Hasegawa, and Hiroo
  Kanamori.
\newblock {Structural heterogeneity in the megathrust zone and mechanism of the
  2011 Tohoku-oki earthquake (Mw 9.0)}.
\newblock \emph{Geophys. Res. Lett.}, 38\penalty0 (17), 2011.

\bibitem[Aochi and Ide(2011)]{Aochi:2011}
Hideo Aochi and Satoshi Ide.
\newblock {Conceptual multi-scale dynamic rupture model for the 2011 off the
  Pacific coast of Tohoku Earthquake}.
\newblock \emph{{Earth Planets Space}}, 63\penalty0 (7):\penalty0 761--765,
  2011.

\bibitem[Nakata et~al.(2011)Nakata, Ando, Hori, and Ide]{Nakata:2011}
Ryoko Nakata, Ryosuke Ando, Takane Hori, and Satoshi Ide.
\newblock {Generation mechanism of slow earthquakes: Numerical analysis based
  on a dynamic model with brittle-ductile mixed fault heterogeneity}.
\newblock \emph{J. Geophys. Res.}, 116\penalty0 (B8):\penalty0 B08308, 2011.

\bibitem[Ando et~al.(2012)Ando, Takeda, and Yamashita]{Ando:2012}
Ryosuke Ando, Naoto Takeda, and Teruo Yamashita.
\newblock {Propagation dynamics of seismic and aseismic slip governed by fault
  heterogeneity and Newtonian rheology}.
\newblock \emph{J. Geophys. Res.}, 117\penalty0 (B11):\penalty0 B11308, 2012.

\bibitem[Skarbek et~al.(2012)Skarbek, Rempel, and Schmidt]{Skarbek:2012}
Robert~M. Skarbek, Alan~W. Rempel, and David~A. Schmidt.
\newblock {Geologic heterogeneity can produce aseismic slip transients}.
\newblock \emph{Geophy. Res. Lett.}, 39\penalty0 (21):\penalty0 L21306, 2012.

\bibitem[Dublanchet et~al.(2013)Dublanchet, Bernard, and
  Favreau]{Dublanchet:2013}
Pierre Dublanchet, Pascal Bernard, and Pascal Favreau.
\newblock {Interactions and triggering in a 3-D rate-and-state asperity model}.
\newblock \emph{J. Geophys. Res.}, 118\penalty0 (5):\penalty0 2225--2245, 2013.

\bibitem[Yabe and Ide(2017)]{Yabe:2017}
Suguru Yabe and Satoshi Ide.
\newblock {Slip-behavior transitions of a heterogeneous linear fault}.
\newblock \emph{J. Geophys. Res.}, 122\penalty0 (1):\penalty0 387--410, 2017.

\bibitem[Fagereng and Diener(2011)]{Fagereng:2011}
\r{A}ke Fagereng and Johann F.~A. Diener.
\newblock {Non-volcanic tremor and discontinuous slab dehydration}.
\newblock \emph{Geophy. Res. Lett.}, 38\penalty0 (15):\penalty0 L15302, 2011.

\bibitem[Fisher and Brantley(2014)]{Fisher:2014}
Donald~M Fisher and Susan~L Brantley.
\newblock {The role of silica redistribution in the evolution of slip
  instabilities along subduction interfaces: Constraints from the Kodiak
  accretionary complex, Alaska}.
\newblock \emph{J. Struct. Geol.}, 69:\penalty0 395--414, 2014.

\bibitem[Audet and B{\"u}rgmann(2014)]{Audet:2014}
Pascal Audet and Roland B{\"u}rgmann.
\newblock {Possible control of subduction zone slow-earthquake periodicity by
  silica enrichment}.
\newblock \emph{Nature}, 510\penalty0 (7505):\penalty0 389--392, 2014.

\bibitem[Taetz et~al.(2018)Taetz, John, Br{\"o}cker, Spandler, and
  Stracke]{Taetz:2018}
Stephan Taetz, Timm John, Michael Br{\"o}cker, Carl Spandler, and Andreas
  Stracke.
\newblock {Fast intraslab fluid-flow events linked to pulses of high pore fluid
  pressure at the subducted plate interface}.
\newblock \emph{Earth Planet. Sci. Lett.}, 482:\penalty0 33--43, 2018.

\bibitem[Rice(1992)]{Rice:1992}
James~R Rice.
\newblock {Fault stress states, pore pressure distributions, and the weakness
  of the San Andreas fault}.
\newblock In Brian Evans and Teng-fong Wong, editors, \emph{Fault Mechanics and
  Transport Properties of Rocks - A Festschrift in Honor of W. F. Brace},
  chapter~20, pages 475--503. Elsevier, 1992.

\bibitem[Skarbek and Rempel(2016)]{Skarbek:2016}
Robert~M. Skarbek and Alan~W. Rempel.
\newblock {Dehydration-induced porosity waves and episodic tremor and slip}.
\newblock \emph{Geochem. Geophys. Geosyst.}, 17\penalty0 (2):\penalty0
  442--469, 2016.

\bibitem[Ide et~al.(2007)Ide, Beroza, Shelly, and Uchide]{Ide:2007}
Satoshi Ide, Gregory~C Beroza, David~R Shelly, and Takahiko Uchide.
\newblock {A scaling law for slow earthquakes}.
\newblock \emph{Nature}, 447\penalty0 (7140):\penalty0 76--79, 2007.

\bibitem[Ben-Zion(2012)]{BenZion:2012}
Yehuda Ben-Zion.
\newblock {Episodic tremor and slip on a frictional interface with critical
  zero weakening in elastic solid}.
\newblock \emph{Geophys. J. Int.}, 189\penalty0 (2):\penalty0 1159, 2012.

\bibitem[Ide(2008)]{Ide:2008}
Satoshi Ide.
\newblock {A Brownian walk model for slow earthquakes}.
\newblock \emph{Geophys. Res. Lett.}, 35\penalty0 (17), 2008.

\bibitem[Ide(2010)]{Ide:2010}
Satoshi Ide.
\newblock {Quantifying the time function of nonvolcanic tremor based on a
  stochastic model}.
\newblock \emph{J. Geophys. Res.}, 115\penalty0 (B8), 2010.

\bibitem[Bostock et~al.(2015)Bostock, Thomas, Savard, Chuang, and
  Rubin]{Bostock:2015}
Michael~G Bostock, Amanda~M Thomas, Genevi\`{e}ve Savard, Lindsay Chuang, and
  Allan~M Rubin.
\newblock {Magnitudes and moment-duration scaling of low-frequency earthquakes
  beneath southern Vancouver Island}.
\newblock \emph{J. Geophys. Res.}, 120\penalty0 (9):\penalty0 6329--6350, 2015.

\bibitem[Gomberg et~al.(2016)Gomberg, Wech, Creager, Obara, and
  Agnew]{Gomberg:2016}
Joan Gomberg, Aaron Wech, Kenneth Creager, Kazushige Obara, and Duncan Agnew.
\newblock {Reconsidering earthquake scaling}.
\newblock \emph{Geophys. Res. Lett.}, 43\penalty0 (12):\penalty0 6243--6251,
  2016.

\bibitem[Kim et~al.(2010)Kim, Clayton, and Jackson]{Kim:2010}
Y~Kim, R~W Clayton, and J~M Jackson.
\newblock {Geometry and seismic properties of the subducting Cocos plate in
  central Mexico}.
\newblock \emph{J. Geophys. Res.}, 115\penalty0 (B6):\penalty0 B06310, 2010.

\end{thebibliography}

%TC:break Acknowledgements and bibliography
\subsection*{Acknowledgements}
We thank Nathalie Cotte for providing the Guerrero GPS time series.
We are grateful to everyone who participates in the maintenance of the GPS network in Guerrero, including Vladimir Kostoglodov, Jorge Real Perez, and Jos\'{e} Antonio Santiago.
We thank Aur\'{e}lien Mordret, Piero Poli, and Nikola\"{i} Shapiro for fruitful discussions.
All observations needed to evaluate the conclusions in the paper are present in the paper and/or the Supplementary Materials.
The Guerrero low-frequency earthquake catalog is available from W.B.F. upon request.
W.B.F. was supported by NSF grant EAR-PF 1452375.
M.C. acknowledges support from the European Research Council (ERC) under the European Union Horizon 2020 research and innovation program (grant agreement No 742335, F-IMAGE).
All authors declare that they have no competing interests.

\clearpage

%:FIGURES
%TC:break Fig. 1

%TC:break Fig. 2

%TC:break Fig. 3

\end{document}